\documentclass{appolb}
\usepackage{cite}
\usepackage{hyperref}
\usepackage{graphicx}
\usepackage{amsmath}
\usepackage[dvipsnames]{xcolor}
\usepackage{animate}
\begin{document}
\title{Equatorial accretion on the Kerr black hole%
\thanks{Talk presented at the 8th Conference of the Polish Society on Relativity (POTOR8), 19-23 September 2022, Banach Center IMPAN, Warsaw, Poland \cite{prezentacja}.}%
}
\author{Patryk Mach \& Andrzej Odrzywo{\l}ek
\address{Instytut Fizyki Teoretycznej, Uniwersytet Jagiello\'{n}ski\\ {\L}ojasiewicza 11, 30-348 Kraków, Poland}
\\
}
\maketitle
\begin{abstract}
We investigate stationary accretion of the collisionless Vlasov gas onto the Kerr black hole, occurring in the equatorial plane. At infinity the gas obeys the Maxwell-J\"{u}ttner distribution, restricted to the equatorial plane. In the vicinity of the black hole, the motion of the gas is governed by the spacetime geometry. We compute accretion rates of the rest-mass, the energy, and the angular momentum, as well as the particle number surface density, focusing on the dependence of these quantities on the asymptotic temperature of the gas and the black hole spin. The accretion slows down the rotation of the black hole. We present preliminary results for a Vlasov gas accretion onto a Kerr black hole moving with a velocity parallel to the equatorial plane.
\end{abstract}

\section{Introduction and motivation}

There is growing evidence for dark matter accretion in the context of supermassive black holes (SMBH) and, in particular, M87* \cite{PhysRevD.77.064023, DeLaurentis_2022}. Recent observations of galaxies at $10\mathop{<}z\mathop{<}14$ \cite{JWST_panic} might require revisiting the basic physics of structure formation in the early Universe ($z>20$), including a possible creation of SMBH's just ($\sim 100$ mln years) after the Big Bang. The latter is probably impossible without the runaway dark matter accretion and/or the existence of primordial black holes \cite{PhysRevLett.126.101104}. It has also been suggested that long-term observations or star trajectories in the vicinity of black holes can provide means of probing the dark matter density profile and probably allow to distinguish between different models \cite{Deviation}. In binary black hole systems such as OJ287 \cite{OJ287}, the drag/lift effects due to dark matter accretion could be also present and perhaps affect the motion of the minor black hole component. For rotating and moving Kerr black holes the dark matter wind accretion might lead to a complicated behaviour involving a drag/lift or a slowdown and the precession of the black hole spin. Results presented in this article for the Vlasov gas accretion onto a Kerr black hole occurring at the equatorial plane provide mandatory intermediate steps towards understanding the full picture, in which the matter is not restricted to the equatorial plane.

Equatorial accretion of the Vlasov gas onto a Kerr black hole has been analysed in \cite{2+1}. In this paper we report preliminary results for an analogous model, in which the (rotating) black hole moves in a direction parallel to the equatorial plane. The gas is assumed to be collisionless, and it is still confined to the equatorial plane of the Kerr spacetime. We obtain stationary solutions for the Vlasov equation, assuming asymptotic conditions corresponding to a two dimensional Maxwell-J\"{u}ttner distribution boosted with a constant velocity along a direction parallel to the equatorial plane. 

\section{Notation, coordinates, and conventions}

We adopt the coordinate system and the notation used in \cite{2+1} for the Kerr black hole at rest. In the case with a moving black hole, or equivalently the Vlasov gas boosted with a constant speed at infinity, our notation follows \cite{PhysRevD.103.024044} to some degree. Details of an efficient implementation of the elliptic function $X$, required for the computation of phase-space integrals, can be found in \cite{Cieslik_2022}. A part of our calculations has been done using \textit{ccgrg ver-2.01} \cite{ccgrg}.

We work in the coordinates $t, r>r_\text{hor}, -\pi<\varphi \le \pi$, where $r_\text{hor}$ corresponds to the location of the outer Kerr horizon. The Kerr metric induced at the equatorial plane reads
$$
\gamma_{\mu \nu}  = \left(
\begin{array}{ccc}
 -1+\frac{2 M}{r} & 0 & -\frac{2 a M}{r} \\
 0 & \frac{r^2}{a^2-2 M r+r^2} & 0 \\
 -\frac{2 a M}{r} & 0 & \frac{2 a^2 M}{r}+a^2+r^2 \\
\end{array}
\right).
$$

The above metric is flat for $M=0$; for $M=0$ and $a=0$ it is explicitly flat in polar coordinates, while for $a \neq 0$ it is explicitly flat in oblate spheroidal coordinates. We have $|\gamma_{\mu \nu}| = |\gamma^{\mu \nu}|^{-1} = -r^2$,
$\gamma_{\mu \nu} \gamma^{\mu \nu}=3$. 

\section{Solution for the Vlasov equation via canonical transformation}

The main goal of this section is to find a general solution of the Vlasov equation
\begin{equation}
\label{Vlasow}
    \frac{\partial f}{\partial x^\mu} \frac{\partial H}{\partial p_\mu}- \frac{\partial f}{\partial p_\nu} \frac{\partial H}{\partial x^\nu}=0,
\end{equation}
where $f(x^\mu, p_\nu) = p(t,r,\varphi, p_t, p_r, p_\varphi)$ denotes the phase-space distribution function. The Hamiltonian corresponding to the time-like geodesic motion reads 
\begin{equation}
\label{H}
    H(t,r,\varphi, p_t, p_r, p_\varphi)=\frac{1}{2} \gamma^{\mu \nu} p_\mu p_\nu = - \frac{1}{2} m^2,
\end{equation}
where $m$ denotes the rest-mass of a gas particle.

The main idea used in \cite{Olivier1, Olivier2} is to introduce a set of new action-angle type variables $(P_\mu,Q^\nu)$, defined by a canonical transformation with the generating function in the form of an action
\begin{equation}
    W = -E t + l_z \varphi + \epsilon_r W_r.
\end{equation}
Here $E = -p_t$ and $l_z = p_\varphi$ are constant, and $\epsilon_r$ is a sign corresponding to the direction of the radial motion. The action $W$ is an integral solution of the Hamilton-Jacobi equation
\begin{equation}
\label{W}
    \gamma^{\mu \nu} \frac{\partial W}{\partial x^\mu} \frac{\partial W}{\partial x^\nu} = -m^2.
\end{equation}
In terms of \textit{old} canonical coordinates $t,r,\varphi$ and \textit{new} canonical momenta $P_0, P_1, P_2$ defined by
$$
P_0 = m, \; P_1 = E,\;  P_2 = l_z,
$$
the generating function $W$ can be written as
$$
W = -P_1 t + P_2 \varphi + \epsilon_r W_r(r, P_0, P_1, P_2).
$$
An explicit expression for $W_r$ can be obtained, but it is not required. Instead, it is sufficient to search for a solution of the following set of equations
\begin{subequations}
\label{canonical_sys}
\begin{eqnarray}
\label{m}
    P_0 & = & m(r, p_t,p_r,p_\varphi) \equiv \sqrt{-\gamma^{\mu \nu} p_\mu p_\nu } = \sqrt{-2 H}, \\
    P_1 & = & -p_t, \\
    P_2 & = & p_\varphi, \\
    Q^0 & = & \partial{W_r} / \partial P_0, \\
    Q^1 & = & - t + \partial{W_r} / \partial P_1, \\
\label{Q2}
    Q^2 & = & \varphi - \partial{W_r} / \partial P_2.
\end{eqnarray}
\end{subequations}

In principle, equations \eqref{canonical_sys} can be solved, yielding the canonical transformation
$( t,r,\varphi, p_t, p_r, p_\varphi ) \longleftrightarrow ( Q^0, Q^1, Q^2, P_0, P_1, P_2 )$. This can be done, using elementary functions, only in the simplest case with $M = 0$, $a = 0$. In practice, we follow a different route. Let us consider $Q^2$, which will appear in the boosted Maxwell-J\"{u}ttner distribution. In integral \eqref{Q2} we change variables acording to
$$
r = M \xi, \quad a= \alpha M, \quad P_1 = \varepsilon P_0, \quad P_2 = \epsilon_\sigma M P_0 (\lambda + \epsilon_\sigma \alpha \varepsilon),
$$
where $\epsilon_\sigma = \pm 1$. This transforms equation \eqref{Q2} into a dimensionless 
form
\begin{equation}
    Q^2 = \varphi - \epsilon_r \epsilon_\sigma X(\xi, \varepsilon, \lambda, \alpha \epsilon_\sigma), 
\end{equation}
where the elliptic function $X$ is a generalization of the one used in the Schwarzschild case,  cf. \cite{Acta}, Eq.~(12). In explicit terms, $X$ is defined by the integral
\begin{equation}
\label{X}
    X(\xi, \varepsilon, \lambda, \alpha) 
    =
    \int_\xi^\infty
    \frac{\left( \lambda + \frac{ \alpha  \varepsilon }{1-\frac{2}{\xi }} \right) d \xi}{\left(\frac{\alpha^2}{1-\frac{2}{\xi }}+\xi ^2\right) \sqrt{
    \varepsilon^2
    -\left(1-\frac{2}{\xi }\right) \left(1+\frac{\lambda
   ^2}{\xi^2}\right)
   -\frac{\alpha^2 +2 \varepsilon  \lambda    \alpha }{\xi^2}  }}.
\end{equation}
The canonical variable $Q^2(t,r,\varphi, p_t, p_r, p_\varphi)$ now reads
\begin{equation}
\label{Q2_explicit}
Q^2 = \varphi - \epsilon_r X\left( \frac{r}{M}, \frac{-p_t}{m(r, p_t,p_r,p_\varphi)}, \epsilon_\sigma \frac{p_\varphi + a p_t}{M m(r, p_t,p_r,p_\varphi)}, \frac{a}{M} \epsilon_\sigma  \right).
\end{equation}
One can verify numerically that the above form satisfies the Vlasov equation \eqref{Vlasow}. The above form allows us to base our calculation on well understood properties and a numerical implementation of the function $X$ \cite{Cieslik_2022,Acta}.

\section{Boosted Maxwell-J\"{u}ttner distribution in canonical variables}

The Maxwell-J\"{u}ttner distribution of a simple gas in the Minkowski spacetime, boosted along the $x$ direction, can be written as
\begin{equation}
\label{fexpr}
f = A \delta(m - m_0) \exp \left[ \frac{\beta}{m_0 \sqrt{1 - v^2}} (p_t - v p_x) \right].
\end{equation}
Here $v$ is the gas velocity, $m_0$ denotes the particle mass, and $\beta = m_0/(k_\mathrm{B}T)$, where $T$ is the temperature, and $k_\mathrm{B}$ denotes the Boltzmann constant. In what follows we set $A=1$. The crucial term is the Cartesian $x$ momentum component, which at the equatorial plane can be expressed in coordinates $(r,\varphi)$ as
\begin{eqnarray*}
p_x & = & p_r \cos{\varphi} - \frac{p_\varphi}{r} \sin{\varphi} \\
& = & -\sqrt{p_r^2 + \left( \frac{p_\varphi}{r} \right)^2 } \sin{[\varphi - \arctan{(p_\varphi/r, p_r})]}.
\end{eqnarray*}
In the second form of the above expression we took advantage of the 2-argument $\arctan$ function, to defer the discussion of the signs of $p_r$ and $p_\varphi$ in four quadrants of the  plane $( p_\varphi/r , p_r )$. Using canonical variables (still for $M = a = 0$), we get
$$
p_x = \sqrt{P_1^2 - P_0^2} \sin{Q^2}, \quad Q^2 = \varphi - \arctan{(p_\varphi/r, p_r)}.
$$

The above formulas provide asymptotic conditions corresponding to the gas moving uniformly at infinity (see, e.g, \cite{PhysRevLett.126.101104}). Expression \eqref{fexpr}, together with the above substitutions, provides a solution for the general Vlasov equation in the equatorial plane of the Kerr spacetime, however $P_0, P_1, Q^2$ have to be replaced by appropriate solutions to \eqref{canonical_sys}. This gives the same form
\begin{equation}
\label{boost}
p_x = \sqrt{P_1^2-P_0^2} \sin{Q^2},
\end{equation}
but now $Q^2$ is given by equation \eqref{Q2_explicit}. While $X \to \arctan$ for $M = a = 0$, there is no two-argument version of the elliptic $X$ function, and the above expression is guaranteed to work only in the first quadrant of the $(r,\varphi)$ plane, i.e., for $0 < \varphi < \pi/2$. For other three quadrants, appropriate phase-jumps must be taken into account. For $v = 0$, all quadrants are identical, while in moving ($v \neq 0$) and non-rotating ($a = 0$) cases the flow has a mirror symmetry. A generic flow around a rotating and moving black hole is different in all quadrants (cf. Fig.~\ref{fig:n_flow}). 

\section{Phase space integrals}

The main goal of this section is to compute the components of the particle current surface density $J_\mu$. An essential part of this calculation consists in evaluating integrals of the form
$$
\iiint 
\begin{bmatrix} 
p_t\\ 
p_r\\ 
p_\varphi\\ 
\end{bmatrix}
e^{\frac{\beta}{m_0 \sqrt{1-v^2}}(p_t - v p_x)} \sqrt{-|\gamma^{\mu \nu}|} \delta(m(r,p_\mu)-m_0) dp_t dp_r dp_\varphi,
$$
where $p_x$ is given by \eqref{boost}. To remove the Dirac delta term and simplify the above expression, we change variables as follows: $( p_t,p_r,p_\varphi ) \to (\bar{m}, \varepsilon, \lambda )$,
$$
\bar{m} = m(r,p_\mu), \quad
\varepsilon = -\frac{p_t}{m(r, p_\mu)}, \quad
\lambda = \frac{\epsilon_\sigma (p_\varphi + a p_t)}{M m(r, p_\mu)},
$$
where the functions $m$ and $H$ are given by equations \eqref{m} and \eqref{H}, respectively. The Jacobian of this transformation reads
$$
\bar{m}^2 M/\sqrt{\varepsilon^2-U_\lambda -{\alpha  (\alpha +2 \varepsilon  \lambda 
   \epsilon_\sigma )}/{\xi ^2}},
$$
where $U_\lambda = (1-2/\xi)(1+\lambda^2/\xi^2)$ is the effective radial potential for Schwarzschild timelike geodesics \cite{Cieslik_2022}.

After the change of variables and the integration over $\bar{m}$, we get
\begin{eqnarray*}
\lefteqn{\frac{m_0^3 M}{\xi} \iint 
\begin{bmatrix} 
-\varepsilon\\ 
-\epsilon_\sigma \frac{\xi^2}{\Delta} \sqrt{\varepsilon^2-U_\lambda -\frac{\alpha  (\alpha +2 \varepsilon  \lambda 
   \epsilon_\sigma )}{\xi ^2} }\\ 
\epsilon_\sigma \lambda M + \varepsilon \alpha M \\ 
\end{bmatrix}} \\
&& \times
\frac{
e^{-\frac{\beta}{\sqrt{1-v^2}} \left[ \varepsilon + \epsilon_\sigma v \sqrt{\varepsilon^2-1} \sin{(\varphi - \epsilon_r \epsilon_\sigma X) } \right]}
}{\sqrt{\varepsilon^2-U_\lambda -\frac{\alpha  (\alpha +2 \varepsilon  \lambda 
   \epsilon_\sigma )}{\xi ^2} }} d\varepsilon d\lambda,
\end{eqnarray*}
where $\Delta = M^2(\xi^2 - 2\xi + \alpha^2)$. The above formula is valid in the first quadrant of the $p_r - p_\varphi$ phase-space. In analogy to Minkowski and Schwarzschild cases, branch cut jumps can be smoothed out by inserting $\pm \pi/2$ terms. Integration limits and the splitting into absorbed/scattered currents are the same as those used in \cite{2+1}, see our equations (68--69). We also denote, in compliance with \cite{2+1}, $\tilde{R} = \varepsilon^2-U_\lambda -{\alpha  (\alpha +2 \varepsilon  \lambda 
   \epsilon_\sigma )}/{\xi ^2}$. Finally, the components of the particle current density can be written as
\iftrue
\begin{subequations}
\label{currents}
\begin{eqnarray}
    J_t^\text{(abs)} & = & -\frac{M m_0^3}{\xi}  \sum_{\epsilon_\sigma=\pm1} \sum_{\epsilon_r=-1} 
    \int_1^\infty \left( \int_0^{\lambda_c} \frac{\varepsilon}{\sqrt{\tilde{R}}} \mathrel{S} 
      d\lambda \right) d \varepsilon,\\
    J_r^\text{(abs)} & = & \frac{M m_0^3 \xi}{\xi(\xi-2) + \alpha^2} \sum_{\epsilon_\sigma=\pm1} \sum_{\epsilon_r=-1} 
    \int_1^\infty \left( \int_0^{\lambda_c} \epsilon_r \mathrel{S} 
      d\lambda \right) d \varepsilon,\\
J_\varphi^\text{(abs)} & = & \frac{M m_0^3}{\xi} \sum_{\epsilon_\sigma=\pm1}  \sum_{\epsilon_r=-1} 
    \int_1^\infty \left( \int_0^{\lambda_c} \epsilon_\sigma \frac{\lambda + \epsilon_\sigma \alpha \varepsilon}{\sqrt{\tilde{R}}} \mathrel{S} 
      d\lambda \right) d \varepsilon,\\
J_t^\text{(scatt)} & = & -\frac{M m_0^3}{\xi} 
\sum_{\epsilon_\sigma=\pm1} \sum_{\epsilon_r=\pm1} 
    \int_{\varepsilon_\text{min}}^\infty \left( \int_{\lambda_c}^{\lambda_\text{max}} \frac{\varepsilon}{\sqrt{\tilde{R}}} \mathrel{S} 
      d\lambda \right) d \varepsilon,\\
J_r^\text{(scatt)} & = & \frac{M m_0^3 \xi}{\xi(\xi-2) + \alpha^2} \sum_{\epsilon_\sigma=\pm1} \sum_{\epsilon_r=\pm1} 
    \int_{\varepsilon_\text{min}}^\infty \left( \int_{\lambda_c}^{\lambda_\text{max}} \epsilon_r \mathrel{S} 
      d\lambda \right) d \varepsilon,\\
    J_\varphi^\text{(scatt)} & = & \frac{M m_0^3}{\xi} \sum_{\epsilon_\sigma=\pm1} \sum_{\epsilon_r=\pm1} 
    \int_{\varepsilon_\text{min}}^\infty 
    \left( 
    \int_{\lambda_c}^{\lambda_\text{max}} \epsilon_\sigma \frac{\lambda + \epsilon_\sigma \alpha \varepsilon}{\sqrt{\tilde{R}}} \mathrel{S}
      d\lambda \right) d \varepsilon,
\end{eqnarray}
\end{subequations}
\fi
where, to shorten notation, we have denoted
$$
S = e^{-\frac{\beta}{\sqrt{1-v^2}} \left\{ \varepsilon +\epsilon_\sigma v \sqrt{\varepsilon^2-1} \sin{\left[ \varphi - \epsilon_r \epsilon_\sigma (-X+\pi/2)  \right]}\right\}}.
$$

\begin{figure}
    \centering
    \animategraphics[controls=all,autoplay,palindrome,height=10cm,every=5,poster=7]{3}{rys/frames/Side_}{1}{400}
    \caption{Flow directions and (surface) particle density $n = \sqrt{-J_\mu J^\mu}$ computed according to equations \eqref{currents}. The asymptotic value of $n$ is denoted as $n_\infty$, and computed using eq.~(50) in \cite{2+1}. Note a highly asymmetric flow for large velocities and Kerr parameters.}
    \label{fig:n_flow}
\end{figure}

Note that $X=X(\xi,\varepsilon,\lambda, \alpha \epsilon_\sigma)$ depends on $\epsilon_\sigma$ via the sign of its last argument, although definition \eqref{X} is chosen in such a way that $X(\xi, \varepsilon,\lambda,0)$ reduces to the form used previously for the moving Schwarzschild black hole \cite{PhysRevLett.126.101104}. Structurally, formulas for $J_\mu^\text{(abs)}$ and $J_\mu^\text{(scatt)}$ are similar, except for the sum over $\epsilon_r$ and integration limits. 

The stress-energy tensor components $T^{\mu \nu}$  can be computed by noticing that integrands of $J_\mu$ and $T_{\mu\nu}$ are similar, cf., e.g., \cite{Cieslik_2020}
$$
\iiint 
\begin{bmatrix} 
p_t^2  & p_t p_r  & p_t p_\varphi\\ 
\cdot  & p_r^2    & p_r p_\varphi\\ 
\cdot  & \cdot& p_\varphi^2\\ 
\end{bmatrix}
e^{\frac{\beta}{m_0 \sqrt{1-v^2}}(p_t - v p_x)} \sqrt{-|\gamma^{\mu \nu}|} \delta(m(r,p_\mu)-m_0) dp_t dp_r dp_\varphi.
$$

Accretion rates can be derived following Sec.\ VI. in \cite{2+1}. For example, the rest-mass accretion rate is
$$
\dot{M} = -2 \pi M m_0 \sum_{\epsilon_\sigma = \pm1} \int_0^\infty \lambda_c(\varepsilon, \alpha,\varepsilon_\sigma) e^{-\frac{\beta  \varepsilon}{\sqrt{1-v^2}}} I_0\left( \beta \gamma v \sqrt{\varepsilon^2-1} \right) \; d \varepsilon,
$$
where $\lambda_c$ can be found in \cite{2+1}, equations (53), and $I_0$ is the modified Bessel function, see \cite{I0}, equation (10.25.2).

\section{Conclusions}

We have investigated a model of the equatorial accretion of dark matter onto a Kerr black hole. Our main findings \cite{2+1} can be summarized as follows: the relation between the mass accretion rate and the black hole rotation parameter exhibits a ``circular'' shape due to shrinkage and expansion of circular photon orbits. The angular momentum accretion rate has a nearly linear dependence on the black hole rotation parameter, except for the values of $\alpha = \pm 1$. The accretion process slows down the black hole rotation, with a timescale of $\tau = c/(24 G \rho_{\infty})$ \cite{prezentacja}.
There are surprising differences between the accretion of a gas confined to the equatorial plane and models in which the motion of the gas is not restricted to the equatorial plane. They are due to the effective dimensionality of these two cases. This fact calls for a construction of a full 3+1 dimensional model of the Vlasov gas accretion in the Kerr spacetime.

\end{document}